\documentclass[12pt]{article}
\usepackage{graphicx}
\usepackage{dcolumn}
\usepackage{bm}
\usepackage{color}
\usepackage{amssymb}
\usepackage{amsmath}
\usepackage{multirow, multicol,times,subfigure, setspace, rotating, url}

\usepackage{algorithm}
\usepackage{algpseudocode}
\usepackage{epstopdf}

\begin{document}

\title{SI-spreading-based network embedding in static and temporal networks}
\author{
Xiu-Xiu Zhan{\small$^{\mbox{1}}$},
Ziyu Li{\small$^{\mbox{1}}$},
Naoki Masuda{\small$^{\mbox{2,3}}$},
Petter Holme{\small$^{\mbox{4}}$},
Huijuan Wang{\small$^{\mbox{1}}$*}}
\maketitle

\vspace{-5mm}

\noindent
$^1$Faculty of Electrical Engineering, Mathematics, and Computer Science,
Delft University of Technology, Mekelweg 4, Delft, The Netherlands, 2628 CD.
$^2$ Department of Mathematics, University at Buffalo, State University of New York, Buffalo,
New York, USA, NY 14260-2900.
$^3$ Computational and Data-Enabled
Science and Engineering Program, University at Buffalo, State University of New York, Buffalo,
New York, USA, NY 14260-2900
$^4$ Tokyo Tech World Research Hub Initiative (WRHI), Institute of Innovative Research,
Tokyo Institute of Technology, Yokohama, Japan, 226-8503.

\begin{abstract} 
Link prediction  can be used to extract missing information, identify spurious interactions as well  as  forecast network evolution. Network embedding is a methodology to assign coordinates to nodes in a low-dimensional vector space. By embedding nodes into vectors, the link prediction problem can be converted into a similarity comparison task. Nodes with similar embedding vectors are more likely to be connected.

Classic network embedding algorithms are random-walk-based. They sample trajectory paths via random walks and generate node pairs from the trajectory paths. The node pair set is further used as the input for a Skip-Gram model, a representative language model that embeds nodes (which are regarded as words) into vectors. In the present study, we propose to replace random walk processes by a spreading process, namely the susceptible-infected (SI) model, to sample paths. Specifically, we propose two SI-spreading-based algorithms, \textit{SINE} and \textit{TSINE}, to embed static and temporal networks, respectively. The performance of our algorithms is evaluated by the missing link prediction task in comparison with state-of-the-art static and temporal network embedding algorithms. Results show that \textit{SINE} and \textit{TSINE} outperform the baselines across all six empirical datasets. We further find that the performance of \textit{SINE} is mostly better than \textit{TSINE}, suggesting that temporal information does not necessarily improve the embedding for missing link prediction. Moreover, we study the effect of the sampling size, quantified as the total length of the trajectory paths, on the performance of the embedding algorithms. The better performance of \textit{SINE} and \textit{TSINE} requires a smaller sampling size in comparison with the baseline algorithms. Hence, SI-spreading-based embedding tends to be more applicable to large-scale networks.
\end{abstract}
\section{Introduction}
\newcommand{\rmnum}[1]{\romannumeral #1}
\newcommand{\Rmnum}[1]{\expandafter\@slowromancap\romannumeral #1@}
\makeatother
Real-world systems can be represented as networks, with nodes representing the components and links representing the connections between them~\cite{newman2003structure, zhang2016dynamics}. The study of complex networks pervades in different fields~\cite{costa2011analyzing}. For example, with biological or chemical networks, scientists study interactions between proteins or chemicals to discover new drugs~\cite{qi2006evaluation,girvan2002community}. With social networks, researchers tend to classify or cluster users into groups or communities, which is useful for many tasks, such as advertising, search and recommendation~\cite{jacob2014learning,traud2012social}. With communication networks, learning the network structure can help understand how information spreads over the networks~\cite{zhang2016dynamics}. These are only a few examples of the important role of analyzing networks. For all these examples, the data may be incomplete. If so, it could be important to be able to predict the link most likely to be missing. If the network is evolving, it could be crucial to forecast the next link to be added. For both of these applications one needs link prediction~\cite{liben2007link, lu2011link, lu2012recommender, martinez2017survey,liu2019computational}.

In link prediction, one estimates the likelihood that two nodes are adjacent to each other based on the observed
network structure~\cite{getoor2005link}. Methods using similarity-based metrics, maximum likelihood algorithms and probabilistic models are major families of link prediction methods~\cite{cui2018survey}.
Recently, network embedding, which embeds nodes into a low-dimensional vector space, has attracted much attention in solving the link prediction problem~\cite{cui2018survey, wang2017community}. The similarity between the embedding vectors of two nodes is used to evaluate whether they would be connected or not.
Different algorithms have been proposed to obtain network embedding vectors. A simplest embedding method is to take the row or column vector in the adjacency matrix, which is called an adjacency vector of the corresponding node,  as the embedding vector. Then, the representation space is $N$-dimensional, where $N$ is the number of nodes. As real-world networks are mostly large and sparse, the adjacency vector of a node is sparse and high-dimensional. In addition, the adjacency matrix only contains the first-order neighborhood information, and therefore the adjacency vector neglects the high-order structure of the network such as paths longer than an edge. These factors limit the precision of network embedding based on the adjacency vector in link prediction tasks. Work in the early 2000s attempted to embed nodes into a low dimension space using dimension reduction techniques~\cite{tenenbaum2000global, roweis2000nonlinear, belkin2002laplacian}. Isomap~\cite{tenenbaum2000global}, locally linear embedding (LLE)~\cite{roweis2000nonlinear} and Laplacian eigenmap~\cite{belkin2002laplacian} are algorithms based on the $k$-nearest graph, where nodes $i$ and $j$ are connected by a link in the $k$-nearest graph if the length of the shortest path between $i$ and $j$ is within the $k$-th shortest among the length of all the shortest paths from $i$ to any other nodes.
 Matrix factorization algorithms decompose the adjacency matrix into the product of two low-dimensional rectangular matrices. The columns of the rectangular matrices are the embedding vectors for nodes. Singular value decomposition (SVD)~\cite{golub1971singular}  is one commonly used and simple matrix factorization. However, the computation complexity of most of the aforementioned algorithms is at least quadratic in terms of $N$, limiting their applicability to large networks with millions of nodes.

Random-walk-based network embedding is a promising family of computationally efficient algorithms. These algorithms exploit truncated random walks to capture the proximity between nodes~\cite{perozzi2014deepwalk,tang2015line,mikolov2013distributed} generally via the following three steps~\cite{grover2016node2vec, cao2018link, zhang2019degree}: (1) Sample the network by running random walks to generate trajectory paths. (2) Generate a node pair set from the trajectory paths: each node on the trajectory path is viewed as a center node, the nearby nodes within a given distance are considered as the neighboring nodes. A node pair in the node pair set is formed by a center node and each of its neighboring nodes.
(3) Apply a word embedding model such as Skip-Gram to learn the embedding vector for each node by using the node pair set as input. Skip-Gram assumes nodes that are similar in topology or content tend to have similar representations~\cite{mikolov2013distributed}. Algorithms have been designed using different random walks to capture high-order structure on networks. For example, DeepWalk~\cite{perozzi2014deepwalk} and Node2Vec~\cite{grover2016node2vec} adopted uniform and biased random walks, respectively, to sample the network structure. In addition, random-walk-based embedding methods have also been developed for temporal networks, signed networks and multilayer networks~\cite{nguyen2018continuous,yuan2017sne, bagavathi2018multi, qu2019temporal}.

In contrast to random-walk-based embedding, here we propose SI-spreading-based network embedding algorithms for static and temporal networks. We deploy the susceptible-infected (SI) spreading process on the given network, either static or temporal, and use the corresponding spreading trajectories to generate the node pair set, which is fed to the Skip-Gram to derive the embedding vectors. The trajectories of an SI spreading process capture the tree-like sub-network centered at the seed node, whereas random walk explores long walks that possibly revisit the same node.
 We evaluate our static network embedding algorithm, which refer to as \textit{SINE}, and temporal network embedding, \textit{TSINE}, via a missing link prediction task in six real-world social networks. We compare our algorithms with state-of-the-art static and temporal network embedding methods. We show that both \textit{SINE} and \textit{TSINE} outperform other static and temporal network embedding algorithms, respectively. In most cases, the static network embedding, \textit{SINE}, performs better than \textit{TSINE}, which additionally uses temporal network information. In addition, we evaluate the efficiency of SI-spreading-based network embedding via exploring the sampling size for the Skip-Gram, quantified as the sum of the length of the trajectory paths, in relation to its performance on the link prediction task. We show that high performance of SI-spreading-based network embedding algorithms requires a significantly smaller sampling size compared to random-walk-based embeddings. We further explore what kind of links can be better predicted to further explain why our proposed algorithms show better performance than the baselines.

The rest of the paper is organized as follows. We propose our method in Section~\ref{Proposed Method}. In Section~\ref{SI based static network sampling}, we propose our SI-spreading-based sampling method for static networks and the generation of the node pair set from the trajectory paths. Skip-Gram model is introduced in Section~\ref{Skip-Gram model}. We introduce an SI-spreading-based sampling method for temporal networks in Section~\ref{SI based temporal network sampling}. In Section~\ref{Evaluation Method}, our embedding algorithms are evaluated on a missing link prediction task on real-world static and temporal social networks.
The paper is concluded in Section~\ref{Conclusions}.

\section{SI-spreading-based Embedding}
\label{Proposed Method}

This section introduces SI-spreading-based network embedding methods. Firstly, we illustrate our SI-spreading-based network embedding method for static networks in Sections~\ref{SI based static network sampling} and~\ref{Skip-Gram model}. Section~\ref{SI based temporal network sampling} generalizes the method to temporal network embedding.

Because we propose the network embedding methods for both static and temporal networks, we start with the notations for temporal networks, of which the static networks are special cases.
A temporal network is represented as $\mathcal{G} = (\mathcal{N}, \mathcal{L})$, where $\mathcal{N}$ is the node set and $\mathcal{L}=\{l(i, j, t), t \in[0, T], i, j\in \mathcal{N}\}$ is the set of time-stamped contacts. The element $l(i, j, t)$ in $\mathcal{L}$ represents a bidirectional contact between nodes $i$ and $j$ at time $t$. We consider discrete time and assume that all contacts have a duration of one discrete time step. We use $[0, T]$ to represent the observation time window, $N = |\mathcal{N}|$ is the number of nodes. The aggregated static network $G=(\mathcal{N}, E)$ is derived from a temporal network $\mathcal{G}$. Two nodes are connected in $G$ if there is at least one contact between them in $\mathcal{G}$. $E$ is the edge set of $G$. The network embedding problem is formulated as follows:

 Given a network $G=(\mathcal{N}, E)$, static network embedding aims to learn a  low-dimensional representation for each node $i \in\mathcal{N}$. The node embedding matrix for all the nodes is given by $\textbf{U}\in R^{d\times N}$, where $d$ is the dimension of the embedding vector ($d < N$). The $i$-th column of $\textbf{U}$, i.e., $\overrightarrow{u_{i}}\in R^{d\times 1}$, represents the embedding vector of node $i$.

\subsection{SI-spreading-based static network sampling}
\label{SI based static network sampling}


The SI spreading process on a static network is defined as follows: each node is in one of the two states at any time step, i.e., susceptible (S) or infected (I); initially, one seed node is infected; an infected node independently infects each of its susceptible neighbors with an infection probability $\beta$ at each time step; the process stops when no node can be infected further. To derive the node pair set as the input for Skip-Gram, we carry out the following steps:
\begin{algorithm}[!ht]
    \caption{Generation of trajectory paths from SI spreading}\label{alg:walkgenerator}
\begin{flushleft}
\textbf{Input:} {$G = (\mathcal{N}, E)$, $B$, $L_{\rm max}$, $\beta$, $m_{i}$ } \\
\textbf{Output:} {node trajectory path set $D$}
\end{flushleft}
    \begin{algorithmic}[1]
    \State Initialize number of context windows $C=0$
    \State Initialize node trajectory path set $D = \varnothing$
\While{$\mathcal{B} - C > 0$}
        \State Randomly choose node $i$ as the seed to start the SI spreading
        \State Generate spreading trajectory tree $\mathcal{T}_{i}(\beta)$
        \State Randomly choose $m_{i}$ trajectory paths $D_{g_{i}} (g_{i}=1, \ldots, m_{i})$ from $\mathcal{T}_{i}(\beta)$
        \For {$g_{i}=1, \ldots, m_{i}$}
        \If{$|D_{g_{i}}| > L_{\rm max}$}
                \State Choose the first $L_{\rm max}$ nodes from $D_{g_{i}}$ to form $D_{g_{i}}^{*}$
                \State Add the trajectory $D_{g_{i}}^{*}$ to $D$
                \State $C = C + |D_{g_{i}}^{*}|$
        \Else{}
              \State Add the trajectory $D_{g_{i}}$ to $D$
              \State $C = C + |D_{g_{i}}|$
        \EndIf
        \EndFor
\EndWhile\label{euclidendwhile}
\State \textbf{return} $D$
\end{algorithmic}
\end{algorithm}
\begin{figure}
\centering
\includegraphics[width=18cm]{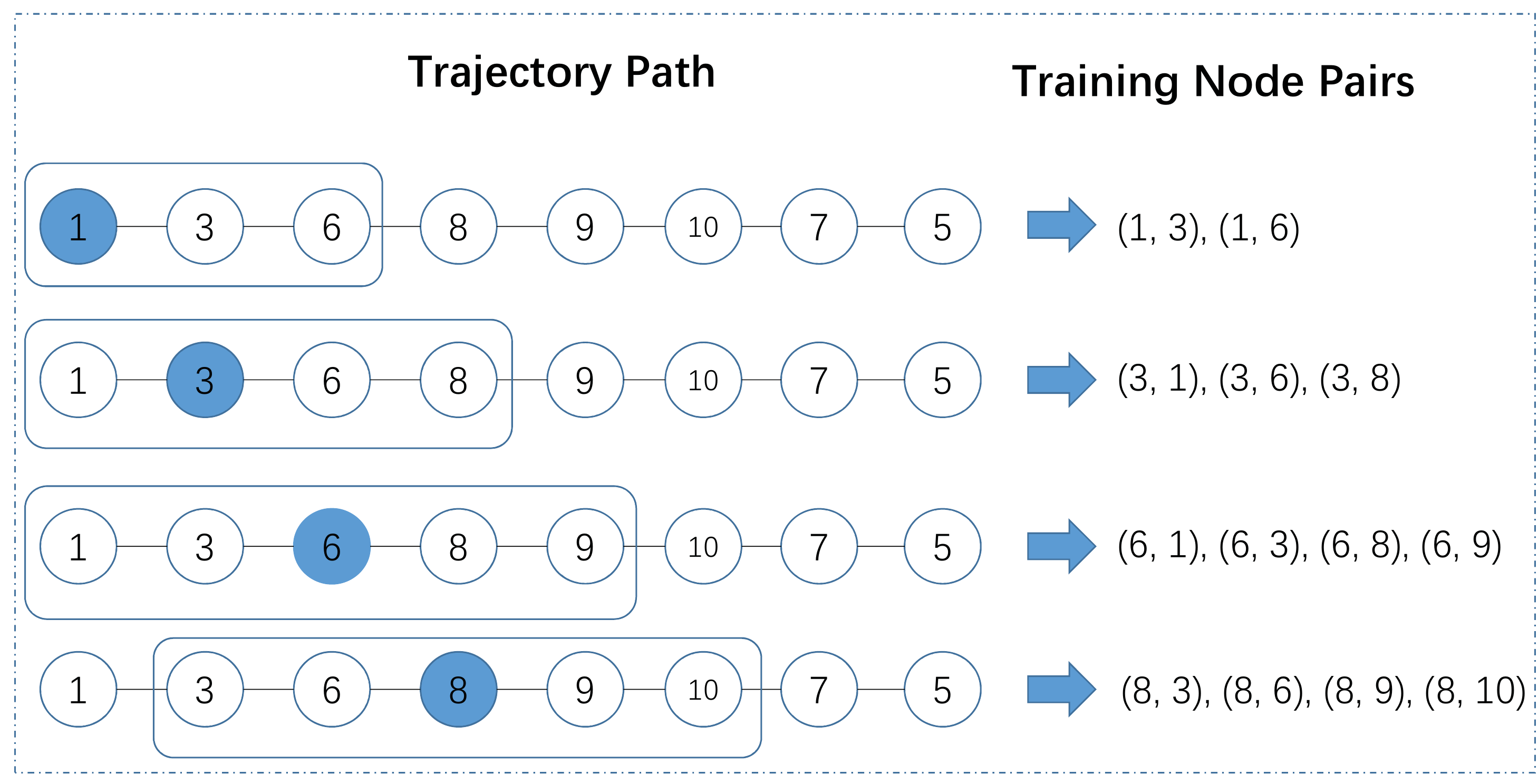}
\caption{\label{Fig:gram}Generating node pairs from a trajectory path $1,3,6,8,9,10,7,5$. The window size $\omega=2$ and only the first four nodes 1, 3, 6 and 8 as the center node are illustrated as examples.}
\end{figure}
\subsubsection{Construction of spreading trajectory paths.} In each run of the SI spreading process, a node $i$ is selected uniformly at random as the seed. The SI spreading process starting from $i$ is performed. The spreading trajectory $\mathcal{T}_{i}(\beta)$ is the union of all the nodes that finally get infected supplied with all the links that have transmitted infection between node pairs.

From each of the spreading trajectory $\mathcal{T}_{i}(\beta)$, we construct $m_{i}$ trajectory paths, each of which is the path between the root node $i$ and a randomly selected leaf node in $\mathcal{T}_{i}(\beta)$. The number $m_{i}$ of trajectory paths to be extracted from $\mathcal{T}_{i}(\beta)$ is assumed to be given by \[m_{i}=\max\left\{1, \frac{\mathcal{K}(i)}{\sum_{j\in\mathcal{N}}\mathcal{K}(j)}m_{\max}\right\},\] where $m_{\max}$ is a control parameter and $\mathcal{K}(i)$ is the degree of the root node $i$ in the static network (or aggregated network).




The trajectory paths may have different lengths (i.e., number of nodes in the path). For a trajectory path whose length is larger than $L_{\max}=20$, we only take the first $L_{\max}$ nodes on the path. For a randomly chosen seed node $i$, we can generate $m_{i}$ trajectory paths from $\mathcal{T}_{i}(\beta)$. We stop running the SI spreading process until the sum of the length of the trajectory paths reaches the sampling size $B=NX$, where $X$ is a control parameter. We consider $X \in \{1,2, 5, 10, 25, 50, 100, 150, 200, 250, 300, 350\}$. We compare different algorithms using the same $B$ for fair comparison~\cite{nguyen2018continuous} to understand the influence of the sampling size. We show how to sample the trajectory paths in Algorithm~\ref{alg:walkgenerator}.




\subsubsection{Node pair set generation.} We illustrate how to generate the node pairs, the input of the Skip-Gram, from a trajectory path in Figure~\ref{Fig:gram}. Consider a trajectory path, $1,3,6,8,9,10,7,5$, starting from node 1 and ending at node 5. We set each node, e.g., node 3, as the center node, and the neighboring nodes of the center node are defined as nodes within $\omega=2$ hops. The neighboring nodes of node 3 are, 1, 6 and 8. We thus obtain ordered node pairs $(3, 1)$, $(3, 6)$, and $(3, 8)$. Thus, we use the union of node pairs centered at each node in each of trajectory path as the input to the Skip-Gram model.


\subsection{Skip-Gram model}
\label{Skip-Gram model}

We illustrate how the Skip-Gram derives the embedding vector for each node based on the input node pair set. We denote by $N_{SI}(i)$ the neighboring set for a node $i$ derived from the SI spreading process. A neighboring node $j$ of $i$ may appear multiple times in $N_{SI}(i)$ if $(i, j)$ appears multiple times in the node pair set.

Let $p(j|i)$ be the probability of observing neighboring node $j$ given node $i$. We model the conditional probability $p(j|i)$ as the softmax unit parametrized by the product of the embedding vectors, i.e., $\overrightarrow{u_{i}}$ and $\overrightarrow{u_{j}}$, as follows:
 \begin{eqnarray}\label{likelihood function}
 p(j|i)
&=& \log \frac{\exp
(\overrightarrow{u_{i}}\cdot \overrightarrow{u_{j}}^{T})}{\sum_{k\in\mathcal{N}}\exp
(\overrightarrow{u_{i}}\cdot \overrightarrow{u_{k}}^{T})}
\end{eqnarray} Skip-Gram is to derive the set of the $N$ embedding vectors that maximizes the log probability of observing every neighboring node from $N_{SI}(i)$ for each $i$. Therefore, one maximizes
\begin{eqnarray}\label{log objective function}
\max\quad \mathcal{O}&=&\sum_{i\in\mathcal{N}}\sum_{j\in N_{SI}(i)}\log p(j|i).
\end{eqnarray}

Equation~(\ref{log objective function}) can be further simplified to
\begin{eqnarray}\label{log objective function1}
\max\quad \mathcal{O} &=&\sum_{i\in\mathcal{N}}\left(-\log Z_{i} + \sum_{j\in N_{SI}(i)}\overrightarrow{u_{i}}\cdot \overrightarrow{u_{k}}^{T}\right),
\end{eqnarray}
where
\begin{equation}
Z_{i}=\sum_{k\in\mathcal{N}} \exp(\overrightarrow{u_{i}}\cdot \overrightarrow{u_{k}}^{T}).
\end{equation}
To compute $Z_{i}$ for a given $i$, we need to traverse the entire node set $\mathcal{N}$, which is computationally costly. To solve this problem, we introduce negative sampling~\cite{mikolov2013distributed}, which randomly selects a certain number of nodes $k$ from $\mathcal{N}$ to approximate $Z_{i}$. To get the embedding vectors for each node, we use the stochastic gradient ascent to optimize  Eq.~(\ref{log objective function1}).

The static network embedding algorithm proposed above from the SI-spreading-based static network sampling and Skip-Gram model is named as \textit{SINE}.
\subsection{SI-spreading-based temporal network sampling}
\label{SI based temporal network sampling}
We generalize \textit{SINE} to the SI-spreading-based temporal network embedding by deploying SI spreading processes on the given temporal network, namely, \textit{TSINE}.
For a temporal network $\mathcal{G} = (\mathcal{N}, \mathcal{L})$, SI spreading follows the time step of the contacts in $\mathcal{G}$. Initially, node $i$ is chosen as the seed of the spreading process. At every time step $t\in [0, T]$, an infected node infects each of its  susceptible neighbor in the snapshot through the contact between them with probability $\beta$. The process stops at time $T$.
We construct the spreading trajectory starting from node $i$ as $\mathcal{T}_{i}(\beta)$, which records the union of nodes that get infected together with the contacts through which these nodes get infected. We propose two protocols to select the seed node of the SI spreading. In the first protocol, we start by selecting uniformly at random a node $i$ as the seed. Then, we select uniformly at random a time step from all the times of contacts made by node $i$ as the starting point of the spreading process, i.e., the time when $i$ gets initially infected. We refer to this protocol as \textit{TSINE1}. In the second protocol, we choose a node $i$ uniformly at random as the seed and start the spreading at the time when node $i$ has the first contact. We refer to this protocol as \textit{TSINE2}.

Both \textit{TSINE1} and \textit{TSINE2} generate
the node pair set from the spreading trajectory $\mathcal{T}_{i}(\beta)$ in the same way as described in Section~\ref{SI based static network sampling}. The node pairs from the node pair set is the input of of Skip-Gram for calculating the embedding vector for each node. The SI-spreading-based temporal network embedding uses the information on the time stamps of contacts in addition to the information used by the static network embedding.

\section{Results}
\label{Evaluation Method}
For the link prediction task in a static network, we remove a certain fraction of links from the given network and predict these missing links based on the remaining links. We apply our static network embedding algorithm to the remaining static network to
derive the embedding vectors for the nodes, which are used for link prediction. For a temporal network, we select a fraction of node pairs that have at least one contact. We remove all the contacts between the selected node pairs from the given temporal network. Then, we attempt to predict whether the selected node pairs have at least one contact or not based on the remaining temporal network.  We use the area under the curve (AUC) score to evaluate the performance of the algorithms on the link prediction task. The AUC quantifies the probability of ranking a random node pair that is connected or has at least a contact higher than a random node pair that is not connected or has no contact.


\subsection{Empirical Networks}
\label{Empirical Networks}
We consider temporal networks, each of which records the contacts and their corresponding time stamps between every node pair. For each temporal network $\mathcal{G}$, one can obtain the corresponding static network $G$ by aggregating the contacts between each node pair over time. In other words, two nodes are connected in static network $G$ if there is at least one contact between them in $\mathcal{G}$. The static network $G$ derived from $\mathcal{G}$ is unweighted by definition. We consider the following temporal social network data sets.
\begin{itemize}
    \item \textit{HT2009}~\cite{isella2011s} is a network of face-to-face contacts between the attendees of the ACM Hypertext 2009 conference.
     \item \textit{Manufacturing Email (ME)}~\cite{michalski2011matching} is an email contact network between employees in a mid-sized manufacturing company.
    \item \textit{Haggle}~\cite{chaintreau2007impact} records the physical contacts between individuals via wireless devices.
    \item \textit{Fb-forum}~\cite{opsahl2013triadic} captures the contacts between students at University of Califonia, Irvine, in a Facebook-like online forum.
    \item \textit{DNC}~\cite{konect:2017:dnc-temporalGraph} is an email contact network in the 2016 Democratic National Committee email leak.
    \item \textit{CollegeMsg}~\cite{opsahl2009clustering} records messages between the users of an online community of students from the University of California, Irvine.
\end{itemize}
Table~\ref{TB:1} provides some properties of the empirical temporal networks. In the first three columns we show the properties of the temporal networks, i.e., the number of nodes ($N$), timestamps ($T$) and contacts ($|\mathcal{L}|$). In the remaining columns, we show the properties of the corresponding aggregate static networks, including the number of links ($|E|$), link density, average degree, and clustering coefficient. The temporal networks are considerably different in size, which ranges from hundreds to thousands of nodes, as well as in the network density and clustering coefficient. Choosing networks with different properties allows us to investigate whether the performance of our algorithms can be consistent across networks.

\begin{table*}[!ht]
\centering
\caption{\label{TB:1}Properties of the empirical temporal networks. The number of nodes ($N$), timestamps ($T$), and contacts ($|\mathcal{L}|$) are shown. In addition, the number of links ($|E|$), link density, average degree, and clustering coefficient of the corresponding static network are shown. }

\begin{tabular}{cccccccccc}
\hline
&Dataset &$N$ &$T$ &$|\mathcal{L}|$ &$|E|$ &Link Density &Average Degree  &Clustering Coefficient\\  \hline
&HT2009 & 113 & 5,246 & 20,818 & 2,196 & 0.3470 & 38.87  &0.5348\\
&ME & 167 & 57,842  & 82,927 & 3,251 & 0.2345 & 38.93  &0.5919\\
&Haggle &274  &15,662   &28,244 &2,124 &0.568 & 15.5  &0.6327\\
&Fb-forum & 899  & 33,515   & 33,720 & 7,046 & 0.0175 & 15.68  &0.0637\\
&DNC & 1,891  &  19,383  & 39,264 & 4,465 & 0.0025 &4.72   &0.2091\\
&CollegeMsg &1,899   &58,911    &59,835  &13,838  & 0.0077 &14.57   &0.1094\\
\hline
\end{tabular}
\end{table*}

\subsection{Baseline algorithms}
\label{Baseline algorithms}
We consider three state-of-the-art network embedding algorithms based on Skip-Gram. These baseline algorithms and the algorithms that we proposed differ only in the method to sample trajectory paths, from which the node pair set, i.e., the input to the Skip-Gram, is derived.
 \textit{DeepWalk}~\cite{perozzi2014deepwalk} and \textit{Node2Vec}~\cite{grover2016node2vec} are static network embedding algorithms based on random walks. \textit{CTDNE}~\cite{nguyen2018continuous} is a temporal network embedding algorithm based on random walks.
\begin{itemize}
\item  \textit{DeepWalk}~\cite{perozzi2014deepwalk} deploys classic random walks on a given static network.

\item  \textit{Node2vec}~\cite{grover2016node2vec} deploys biased random walks on a given static network. The biased random walk gives a trade-off between breadth-first-like sampling and depth-first-like sampling of the neighborhood, which is controlled via two hyper-parameters $p$ and $q$. We use a grid search over $p, q \in \{0.01, 0.25, 0.5, 1, 2, 4\}$ to obtain embeddings that achieve the largest AUC value for link prediction.

\item  \textit{CTDNE}~\cite{nguyen2018continuous}: \textit{CTDNE} is a temporal network embedding algorithm based on temporal random walks. The main idea is that the timestamp of the next temporal contact on the walk should be larger than the timestamps of previously traversed contacts.
Given a temporal network $\mathcal{G} = (\mathcal{N}, \mathcal{L})$, the starting contact for the temporal random walk is selected uniformly at random. Thus, every contact has probability $1/|\mathcal{L}|$ to be selected as the starting contact. Assume that a random walker visits node $i$ at time step $t$. We define $\Gamma_{t}(i)$ as the set of nodes that have contacted node $i$ after time $t$ allowing duplicated elements. A node may appear multiple times in $\Gamma_{t}(i)$ because it may have multiple contacts with node $i$ over the course of time. The next node to walk to is uniformly selected from $\Gamma_{t}(i)$, i.e., every node in $\Gamma_{t}(i)$ is chosen with probability $1/|\Gamma_{t}(i)|$. Nguyen et al.~\cite{nguyen2018continuous} generalized the starting contact and the successor node of a temporal walk to other distributions beyond the uniform distribution illustrated here.
 When we compare the performance of the algorithms on link prediction, we explore the embeddings that give the largest AUC value for link prediction of \textit{CTDNE} by taking into account all possible generalizations proposed by Nguyen et al.
 \end{itemize}

In our SI-spreading-based algorithms for both static and temporal networks, we set $\beta \in \{0.001, 0.01, 0.1, \linebreak 0.2, 0.3, 0.4, 0.5,  0.6,  0.7,0.8, 0.9, 1.0\}$. We use $\omega=10$ and embedding dimension $d=128$ for our algorithms and the baseline algorithms.

\subsection{Performance Evaluation}
\label{Performance Evaluation}

\subsubsection{Training and test sets}
In this section, we illustrate how to generate the training and test sets in the link prediction task  in temporal and static networks.
We run the network embedding algorithms on the corresponding training set and obtain embedding vector for each node,
and use the AUC to evaluate the link prediction performance in the test set.

 Given a temporal network $\mathcal{G}$, we select uniformly at random $75\%$ node pairs among the node pairs that have at least one contact between them in $\mathcal{G}$ as the training set for temporal embedding algorithms, including all the contacts and their timestamps. The training set for static network embedding algorithms is the aggregation of the training set for temporal embedding algorithms. In other words, for every node pair, there is a link between the two nodes in the training set for static network embedding if and only if they have at least one contact in the training set for temporal embedding algorithms.

 We use the remaining $25\%$ node pairs among the node pairs that have at least one contact of $\mathcal{G}$ as the positive links in the test set. We label these node pairs 1. Then, we uniformly randomly sample an equal number of node pairs in $\mathcal{G}$ which have no contact between them. These node pairs are used as negative links in the test set, which we label 0. The same test set is used for the link prediction task in both temporal and static networks.

For each temporal network data set, we randomly split the network to obtain the training and test set according to the procedures given above five times. Both random walks and SI spreading processes are stochastic.  For each split data, we run each algorithm on the training set and perform the link prediction on the test set for ten realizations. Therefore, we obtain ten AUC scores for each splitting of the data into the training and test sets, evening the randomness stemming from stochasticity of the random walk or SI spreading processes. We obtain the AUC score for each algorithm with a given parameter set as an average over 50 realizations in total.

\subsubsection{Evaluation Results}

\begin{table}[!ht]
\centering
\caption{\label{TB:AUC}AUC scores for link prediction. All the results shown are the average over 50 realizations. Bold indicates the optimal AUC among the embedding algorithms, $^{*}$ indicates the optimal AUC among all the algorithms. L2, L3, L4 are the short for link prediction metrics which counts the number of $l=2,3,4$ paths, respectively.}
\resizebox{\textwidth}{18mm}{
\begin{tabular}{ccccccccccc}
        \hline
           &Dataset &DeepWalk &Node2Vec &CTDNE &TSINE1 &TSINE2 & SINE & L2 & L3 & L4  \\ \hline
        &HT2009 & 0.5209 & 0.5572 & 0.6038 & 0.6740 &\textbf{0.6819} &0.6726  & $0.7069^{*}$  & 0.7066 & 0.7055 \\
        &ME & 0.6439 & 0.6619  & 0.6575 & 0.7329 &0.7462 &\textbf{0.7744} & 0.7855 & $0.7878^{*}$ & 0.7790 \\
        &Haggle & 0.3823  & 0.7807   & 0.7796 & 0.8051 &0.8151 &$\textbf{0.8267}^{*}$   & 0.8167 & 0.8255 &0.8226 \\
        &Fb-forum &0.5392   &0.6882    & 0.6942 & 0.7104 &0.7195 &$\textbf{0.7302}^{*}$  & 0.5606 & 0.7179 &0.7203 \\
         &DNC &0.5822   &0.5933 &0.7274 & 0.7539 &0.7529 &\textbf{0.7642} &$0.7704^{*}$   &0.7627  &0.7193\\
         &CollegeMsg &0.5356 &0.5454 &0.7872  & 0.8257 &0.8321 &\textbf{0.8368} &0.7176  &$0.8609^{*}$  &0.8203\\
        \hline
\end{tabular}}
\end{table}
\begin{figure*}[!ht]
\centering
	\includegraphics[width=7.5cm]{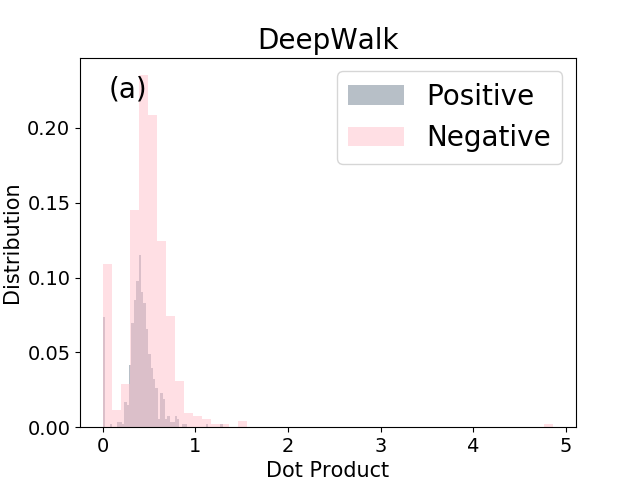}
	\includegraphics[width=7.5cm]{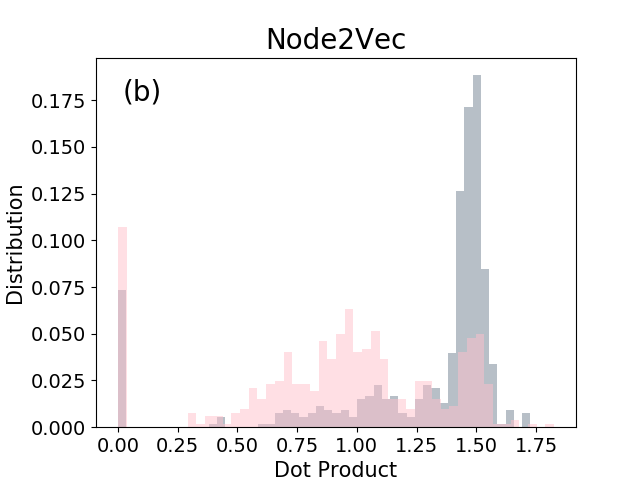}
	\includegraphics[width=7.5cm]{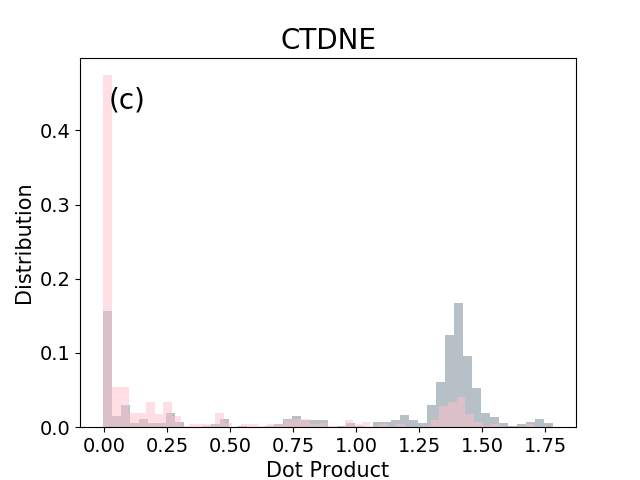}
	\includegraphics[width=7.5cm]{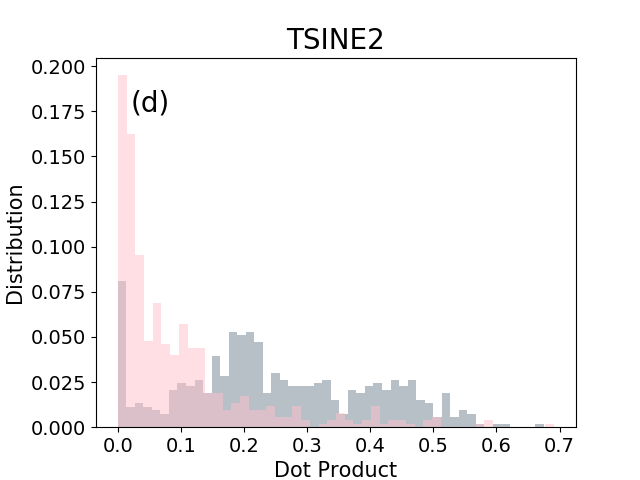}
	\includegraphics[width=7.5cm]{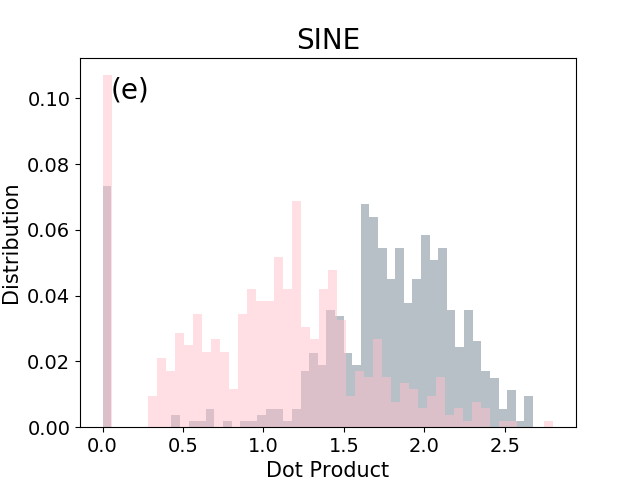}
    \caption{The dot product distribution of the two end nodes' embedding vectors of the positive and negative links in the test set. We show the result of the \textit{Haggle} data set. For each algorithm, we use the same parameter settings as that of Table~\ref{TB:AUC} to obtain the embeddings. Dot products of positive links are shown in grey. Negative links are shown in pink. The results are shown for algorithms (a) \textit{DeepWalk}; (b) \textit{Node2Vec}; (c) \textit{CTDNE}; (d) \textit{TSINE2} and (e) \textit{SINE}.}
    \label{fig:dot_product_hist_Haggle}
\end{figure*}
We summarize the overall performance of the algorithms on missing link prediction in Table~\ref{TB:AUC}.
For each algorithm, we tune the parameters and show the optimal average AUC score. Among the static network embedding algorithms, \textit{SINE} significantly outperforms \textit{DeepWalk} and \textit{Node2Vec}. The improvement in the AUC score is up to 30\% on the \textit{CollegeMsg} dataset. Embedding algorithms \textit{CTDNE}, \textit{TSINE1} and \textit{TSINE2} are for temporal networks. The SI-spreading-based algorithms (i.e., \textit{TSINE1} and \textit{TSINE2}) also show better performance than random-walk-based one (\textit{CTDNE}). Additionally, \textit{TSINE2} is slightly better than \textit{TSINE1} on all data sets. Therefore, we will focus on \textit{TSINE2} in the following analysis. In fact, \textit{SINE} shows better performance than temporal network embedding methods including \textit{TSINE2} on all data sets except for \textit{HT2009}.
It has been shown that temporal information is important for learning embeddings~\cite{nguyen2018continuous, zuo2018embedding, zhou2018dynamic}. However, up to our numerical efforts, \textit{SINE} outperforms the temporal network algorithms although \text{SINE} deliberately neglects temporal information.

\begin{figure*}[!ht]
\centering
	\includegraphics[width=16cm]{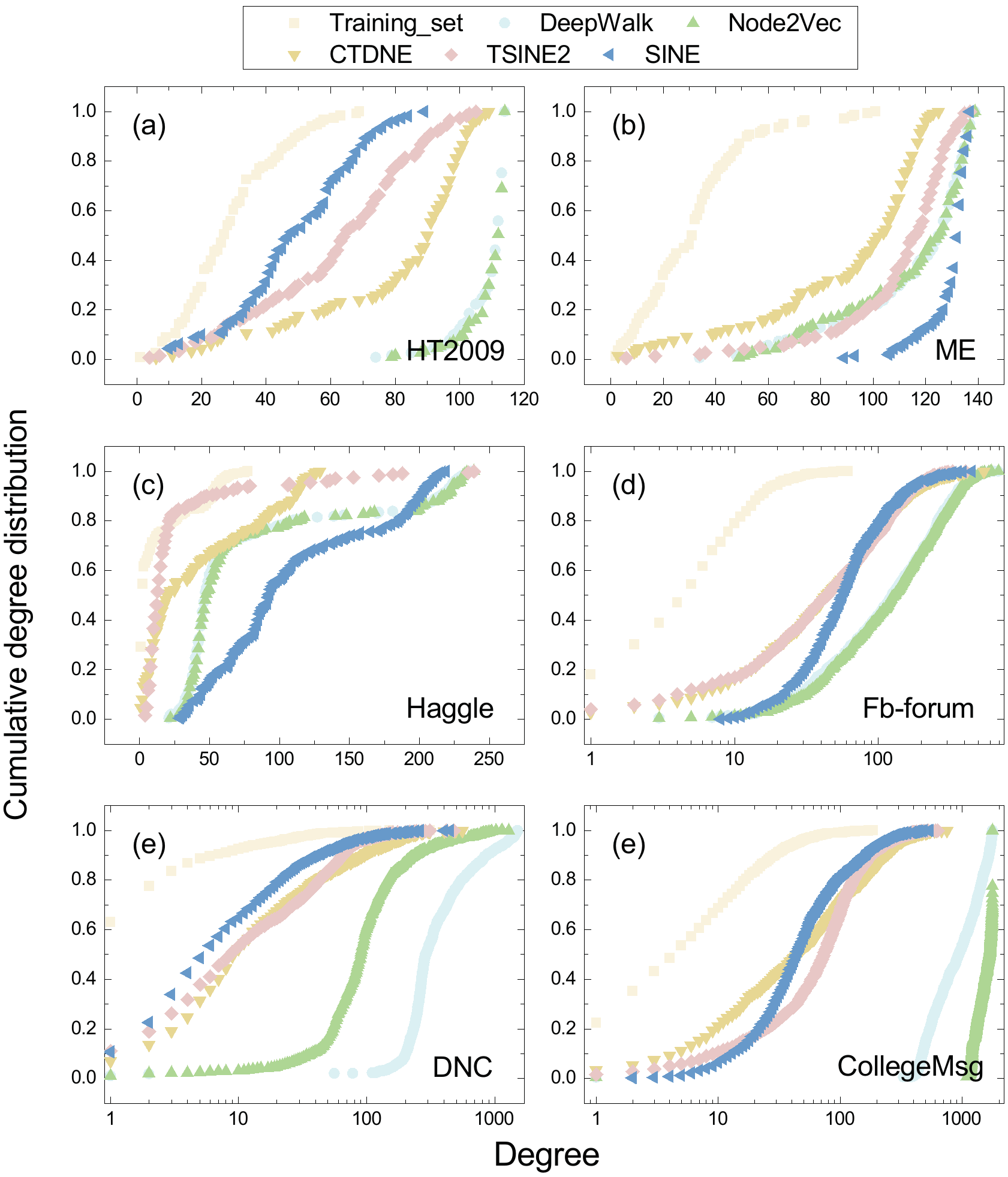}
    \caption{Cumulative degree distribution of the static network derived from the training set and that of the sampled networks $G_{S}$ from different algorithms. We show the results for (a)\textit{HT2009}; (b)\textit{ME}; (c)\textit{Haggle}; (d)\textit{Fb-forum}; (e)\textit{DNC}; (f)\textit{CollegeMsg}.}
    \label{fig:degree distri}
\end{figure*}

To get insights into the different performance among the embedding algorithms, we further investigate the distribution of the dot product of node embedding vectors. Given a link $(i, j)$ in the test set, we compute the dot product of the two end nodes' embedding  vectors, i.e., $\overrightarrow{u_{i}}\cdot \overrightarrow{u_{j}}^{T}$. We show the dot product distribution for the positive links and negative links in the test set separately. For each embedding algorithm, we consider only the parameter set that maximizes the AUC, i.e., the parameter values with which the results are shown in Table~\ref{TB:AUC}. We show the distribution of the dot product for \textit{Haggle} in Figure~\ref{fig:dot_product_hist_Haggle} and for the other data sets in  Figure~S1--S5 in the Appendix.
Compared to the random-walk-based algorithms, \textit{TSINE2} and \textit{SINE} yield more distinguishable distributions between the positive (grey) and the negative links (pink). This result supports the better performance of SI-spreading-based embeddings than random-walk-based ones.

The embedding algorithms differ only in the sampling method to generate the node pair set. These algorithms use the same Skip-Gram architecture, which takes the node pair set as input, to deduce the embedding vector for each node. We explore further how the algorithms differ in the node pair sets that they sampled. The objective is to discover the relation between the properties of the sampled node pairs and the performance of an embedding method.
We represent the node pair set generated by an embedding method as a network $G_{S}=(\mathcal{N}, E_{S})$, so called the sampled network. Two nodes are connected in $G_{S}$ if they form a node pair in the node pair set. It should be noted that $G_{S}$ is an unweighted network. For each algorithm, with the parameter set that maximizes the AUC, we show the cumulative degree distribution of its sampled network $G_{S}$ in Figure~\ref{fig:degree distri}. The cumulative degree distribution of the training set for static network is also given. Compared to the cumulative degree distribution of the training set, the sampled networks tend to have a higher node degree. Zhang et al.\ and Gao et al.~\cite{zhang2019degree,gao2018bine} have shown that when the degree distribution of $G_{S}$ is closer to that of the training set, the prediction performance of a random-walk-based algorithm tends to be better. Even though SI-spreading based algorithms perform the best across the data sets, we have not found a direct relation  between the performance of the embedding algorithm and similarity between the degree distribution of the sampled network and that of the training set.
\begin{figure*}[!ht]
\centering
	\includegraphics[width=18cm]{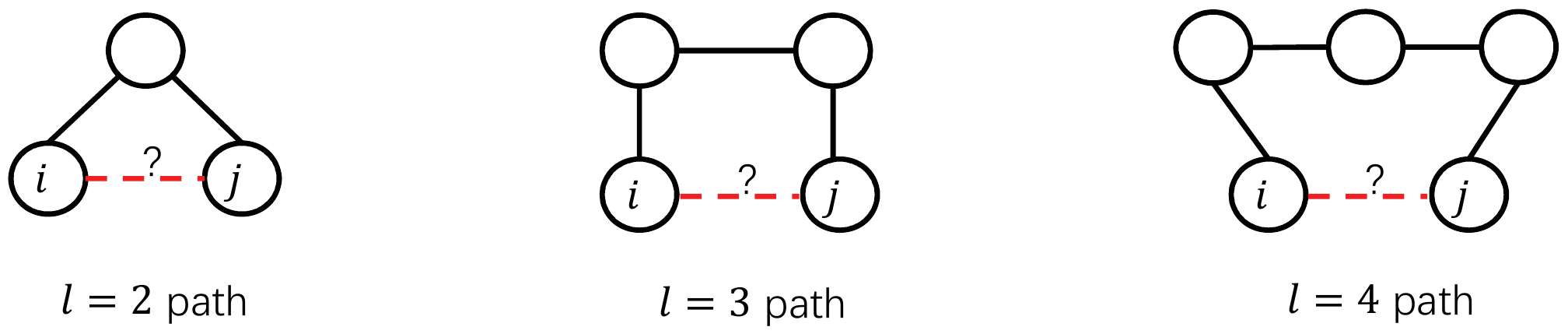}
    \caption{Illustration of $l$ paths between a pair of nodes $i$ and $j$.
    Here we show $l=2,3,4$.}
    \label{fig:l-path}
\end{figure*}

Similarity-based methods such as the number of $l=2,3,4$ paths  have been used for link prediction problem~\cite{lu2011link}. A $l$ path between two nodes refers to a path  that contains $l$ links. We show examples of $l=2, 3, 4$ path between a node pair $i$ and $j$ in Figure~\ref{fig:l-path}.
Kov\'{a}cs et al.~\cite{kovacs2019network} have shown that $l$ paths ($l=3, 4$) outperform existing link prediction methods in predicting protein interaction. Cao et al.~\cite{cao2019network} found that network embedding algorithms based on random walks sometimes perform worse in link prediction than the number of $l=2$ paths or equivalently the number of common neighbors. This result suggests a limit of random-walk-based embedding in identifying the links between node pairs that have many common neighbors.
Therefore, we explore further whether our SI-spreading-based algorithms can overcome this limitation, thus possibly explain their outperformance.

\begin{figure*}[!ht]
\centering
	\includegraphics[width=18cm]{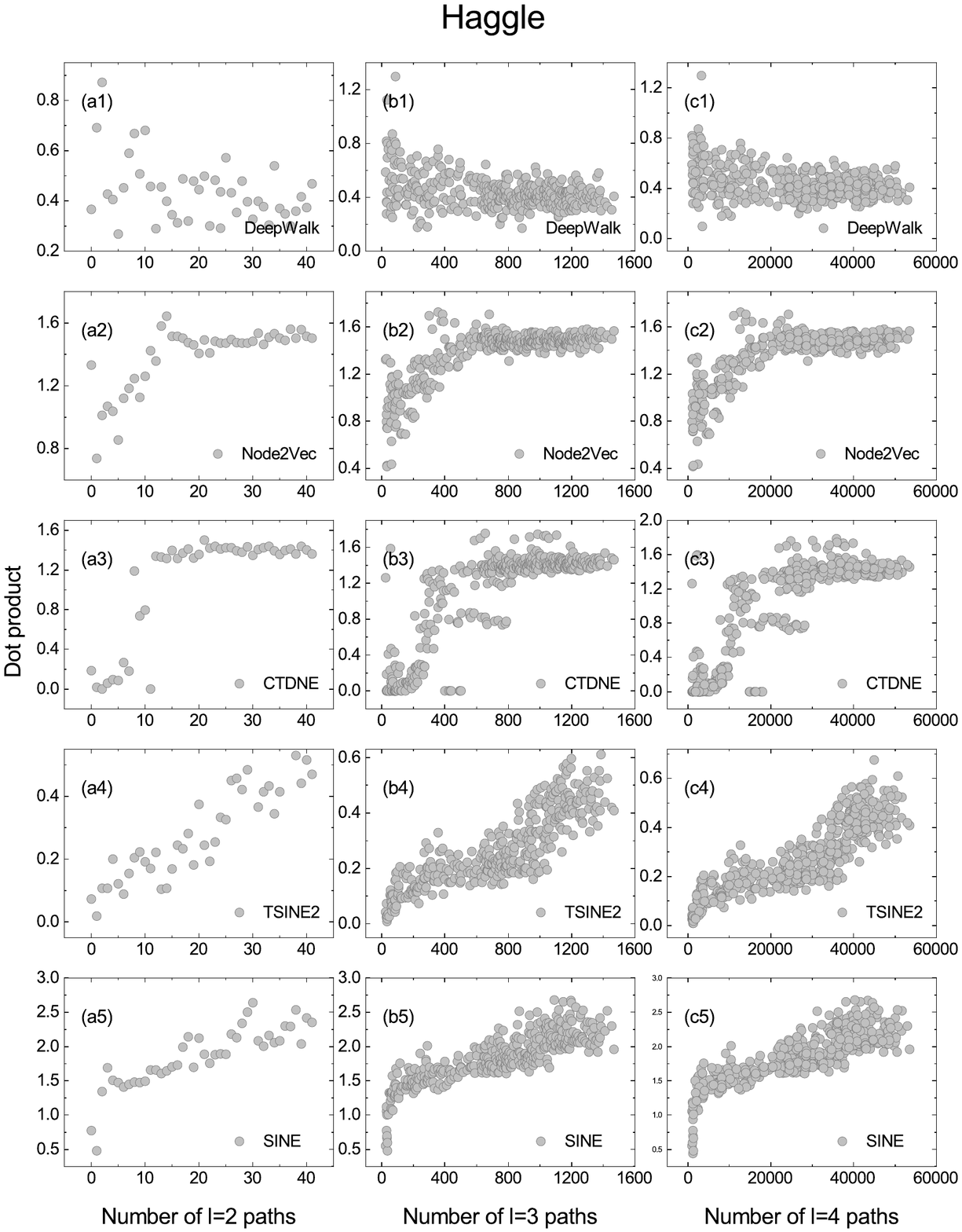}
    \caption{Relation between the dot product of the two end nodes' embedding vectors and the number of $l=2,3,4$ paths between the two end nodes of the positive links in the test set for \textit{Haggle} data set. (a1--a5), (b1--b5) and (c1--c5) are the results for the number of $l=2,3,4$ paths, respectively. }
    \label{fig:Haggle-path-len2-dot-product}
\end{figure*}

 We investigate what kind of network structure surrounding links makes them more easily be predicted.
 For every positive link in the test set, we study its two end nodes' topological properties (i.e., the number of $l=2$, $l=3$ and $l=4$ paths) and the dot product of the embedding vectors of its two end nodes. Given a network, the parameters of each embedding algorithm are tuned to maximize the AUC, as given in Table 2.  We take the data set \textit{Haggle} as an example. Figure~\ref{fig:Haggle-path-len2-dot-product} show the relation between the dot product of the embedding vectors and the number of $l=2,3,4$ paths of the two end nodes of a positive link in the test set for all the embedding methods. The Pearson correlation coefficient (PCC) between the two variables for all the networks and algorithms is given in Table~S1 in the Appendix. Figure~\ref{fig:Haggle-path-len2-dot-product} and Table~S1 together show that the dot product of the embedding vectors constructed from \textit{TSINE2} and \textit{SINE} is more strongly correlated with the number of $l$ paths, where $l=2$, 3 or 4, than the random-walk-based embeddings. This result suggests that SI-spreading-based algorithms may better predict the links whose two end nodes have many $l$-paths, thus overcoming the limit of random-walk-based embedding algorithms.



 The number of $l=2,3$ paths has been used to predict links in ~\cite{lu2011link,kovacs2019network,cao2019network}. The observation and the limit of random-walk-based embedding algorithms motivate us to use the number of $l=2,3,4$ paths between a node pair to predict the missing links.
 Take $l=2$ paths as an example. For every link in the test set, the number of $l=2$ paths between the two end nodes in the training set is used to estimate the likelihood of connection between them.  In the networks we considered, two end nodes of a link tend to be connected by $l=2$, $l=3$ and $l=4$ paths (see Figures~\ref{fig:Haggle-path-len2-dot-product}). Table~\ref{TB:AUC} ($L2,L3,L4$ shown in the table correspond to the method of using the number of $l=2,3,4$ path for link prediction) shows that in such networks, the similarity-based methods do not evidently outperform the SI-spreading-based embedding. Actually, the SI-spreading-based embedding performs better in two out of six networks.

\begin{figure}[!ht]
\centering
\includegraphics[width=18cm]{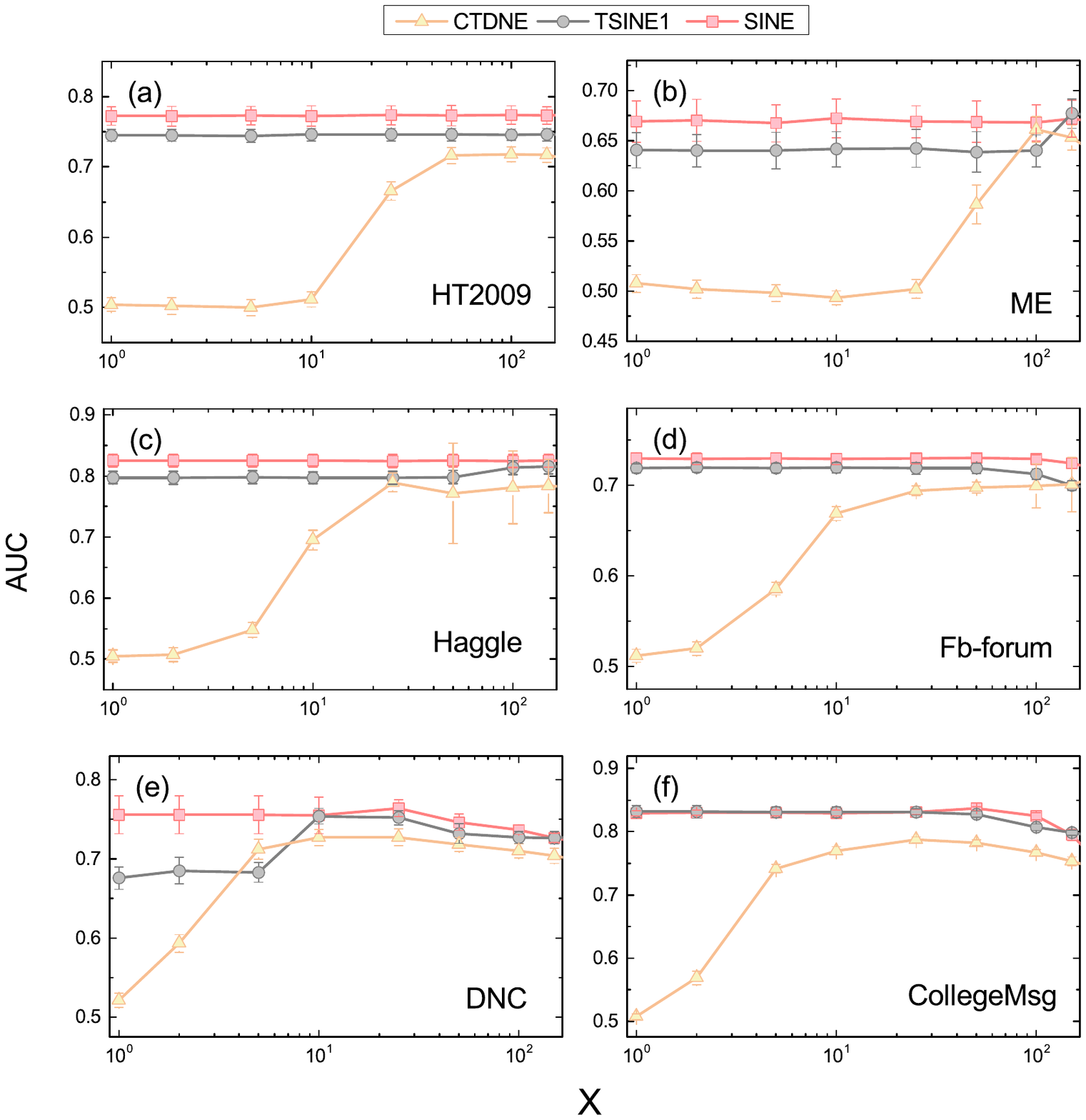}
    \caption{\label{Fig:X} Influence of the sampling size $B=NX$ on the link prediction performance, i.e., AUC score. The error bar shows the standard deviation of the AUC score calculated on the basis of 50 realizations.
    We show the results for (a)\textit{HT2009}; (b)\textit{ME}; (c)\textit{Haggle}; (d)\textit{Fb-forum}; (e)\textit{DNC}; (f)\textit{CollegeMsg}.}
\end{figure}

Next, we study the effect of the sampling size, $B$, on the performance of each algorithm. The sampling size is quantified as the the total length of the trajectory paths as defined in Section~\ref{SI based static network sampling}. Given a network, we set $\mathcal{B}=NX$, where $N$ is the size of the network and $X \in \{1,2, 5, 10, 25, 50, 100, 150\}$. We evaluate our SI-spreading-based embedding algorithms \textit{SINE} and \textit{TSINE2}, and one random-walk-based embedding algorithm \textit{CTDNE}, because \textit{CTDNE} performs mostly the best among all random-walk-based algorithms. The result is shown in Figure~\ref{Fig:X}. For each $X$, we tune the other parameters to show the optimal AUC in the figure.  Both \textit{SINE} and \textit{TSINE2} perform better than \textit{CTDNE} and are relatively insensitive to the sampling size. This means that they achieve a good performance even when the sampling size is small, even with $X=1$. The random-walk-based algorithm, \textit{CTDNE}, however, requires a relatively large sampling size to achieve a comparable performance with \textit{SINE} and \textit{TSINE2}.

\begin{figure}
\centering
\includegraphics[width=18cm]{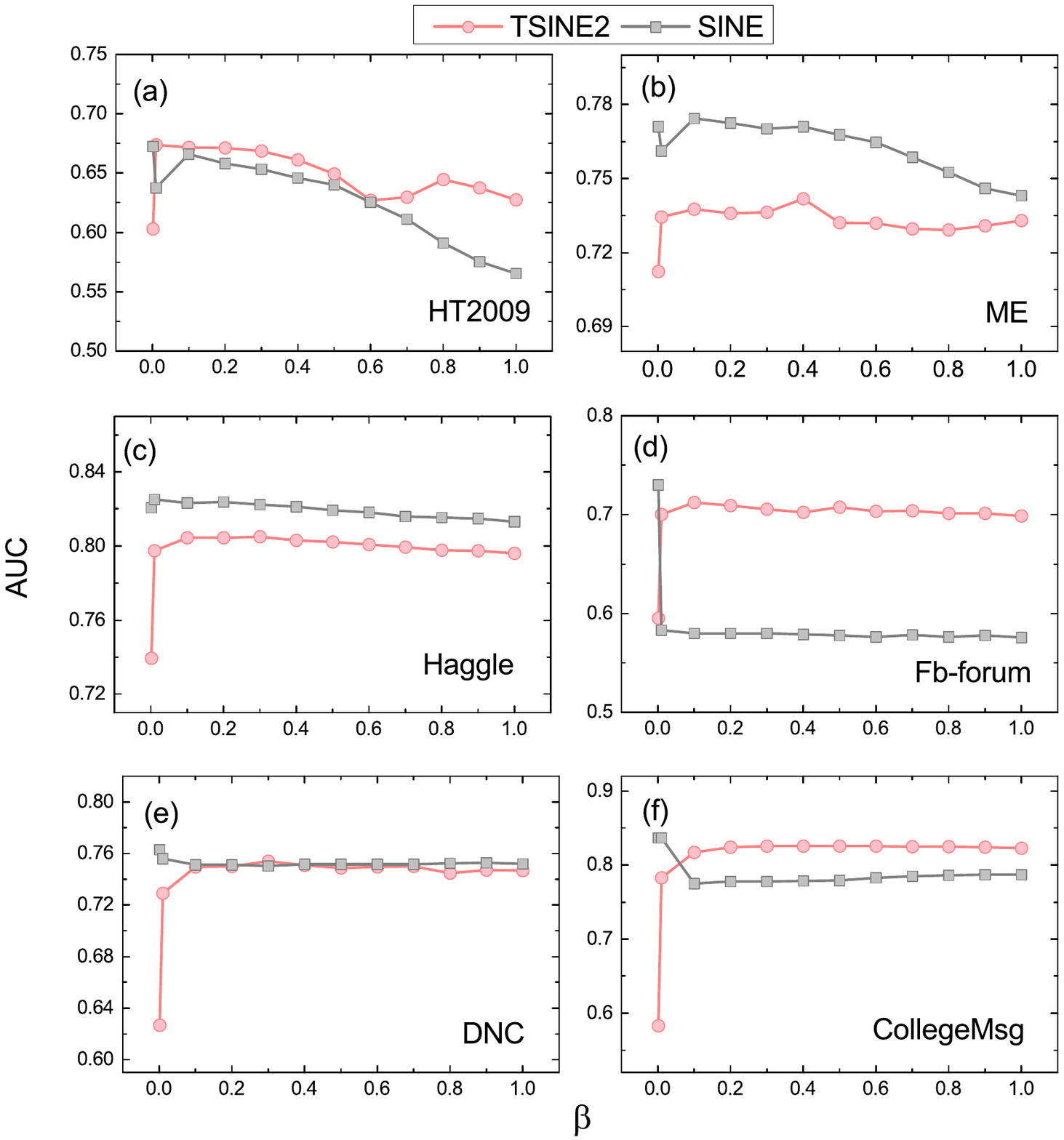}
    \caption{\label{Fig:AUC-beta-opt} AUC as a function of $\beta$. We show the results for (a)\textit{HT2009}; (b)\textit{ME}; (c)\textit{Haggle}; (d)\textit{Fb-forum}; (e)\textit{DNC}; (f)\textit{CollegeMsg}.}
\end{figure}

Finally, the AUC as a function of the infection probability, $\beta$, is shown in Figure~\ref{Fig:AUC-beta-opt}. For each $\beta$, we tune the other parameters to show the optimal AUC. The SI-spreading-based algorithms achieve high performance with a small infection probability ($0.001 \leq \beta \leq 0.1$) for all the data sets. The high performance of SI-spreading-based embedding algorithms with  the small value of $X$ and $\beta$ across different networks motivates the further study
whether one can optimize the performance by searching a smaller range of the parameter values.

\section{Conclusions}
\label{Conclusions}
In this paper, we proposed network embedding algorithms based on SI spreading processes in contrast to the previously proposed embedding algorithms based on random walks~\cite{zhan2019information, zhan2018coupling}. We further evaluated the embedding algorithms on the missing link prediction task.
 The key point of an embedding algorithm is how to design a strategy to sample trajectories to obtain embedding vectors for nodes. We used the SI model to this end. The algorithms that we proposed are \textit{SINE} and \textit{TSINE}, which use static and temporal networks, respectively.


On six empirical data sets, the SI-spreading-based network embedding algorithm on the static network, i.e., \textit{SINE}, gains much more improvement than state-of-the-art random-walk-based network embedding algorithms across all the data sets. The SI-spreading-based network embedding algorithms on the temporal network, \textit{TSINE1} and \textit{TSINE2}, also show better performance than the temporal random-walk-based algorithm. Temporal information provides additional information that may be useful for constructing embedding vectors~\cite{nguyen2018continuous, zuo2018embedding, zhou2018dynamic}. However, we find that \textit{SINE} outperforms \textit{TSINE}, which uses timestamps of the contacts. This result suggests that temporal information does not necessarily improve the embedding for missing link prediction. Moreover, when the sampling size of the Skip-Gram is small, the performance of the SI-spreading-based embedding algorithms is still high. Sampling trajectory paths takes time especially for large-scale networks. Therefore, our observation that the SI-spreading-based algorithms require less samples than other algorithms promises the applicability of the SI-spreading-based algorithms to larger networks than the random-walk-based algorithms. Finally,  we show insights of why SI-spreading-based embedding algorithms performs the best by investigating what kind of links are likely to be predicted.

We deem that the following future work as important. We have already applied susceptible-infected-susceptible (SIS) model and evaluated the SIS-spreading-based embedding. However, this generalization has not improved the performance in the link prediction task. Therefore, one may explore whether or not sampling the network information via the other spreading processes, such as susceptible-infected-recovered (SIR) model, further improves the embedding. It is also interesting to explore further the performance of the SI-spreading-based algorithms in other tasks such as classification and visualization. Moreover, the SI-spreading-based sampling strategies can also be generalized to other types of networks, e.g., directed networks, signed networks, and multilayer networks.

\section{Competing interests}
  The authors declare that they have no competing interests.

\section{Author's contributions}
All authors planed the study; X.Z. and Z.L. performed the experiments, analyzed the data and prepared
the figures. All authors wrote the manuscript.

\section{Acknowledgements}
We thank the SocioPatterns collaboration (http://
www.sociopatterns.org) for providing the data sets. This work has been partially supported by the China Scholarship Council (CSC).
\clearpage

\bibliographystyle{naturemag} 
\bibliography{SI-embedding-arxiv-version}      





\end{document}